\documentclass[final,authoryear,5p,times,twocolumn]{elsarticle}

\usepackage{graphics}

\usepackage{amssymb}

\journal{no journal}

\begin{document}

\begin{frontmatter}



\title{Bacterial survival in Martian conditions}


\author{Giuseppe Galletta\corref{gg}$^{a,b}$, Giulio Bertoloni$^c$  and Maurizio D'Alessandro$^d$}

\cortext[gg]{Corresponding author. \\
E-mail address: giuseppe.galletta@unipd.it}

\address{$^a$Dipartimento di Astronomia, Universit\`a di Padova, vicolo Osservatorio 3, 
35122 Padova, Italy \\ $^b$ CISAS "G. Colombo", Universit\`a di Padova, via Venezia 15, 35131 Padova, Italy \\
$^c$Dipartimento di Istologia, Microbiologia e Biotecnologie Mediche, Universit\`a di Padova,  via Gabelli, 35121 Padova, Italy \\
$^d$INAF - Osservatorio Astronomico di Padova,  vicolo Osservatorio 5, 35122 Padova, Italy}

\begin{abstract}

Terrestrial lifeforms show a strong capability to adapt to very harsh environments and to survive even to strong shocks as those derived from meteorite impacts. These findings increase the possibility to discover traces of life on a planet like Mars, that had past conditions similar to the early Earth but now is similar to a very cold desert, irradiated by intense solar UV light. 

We shortly discuss the observable consequences of the two hypotheses about the origin of life on Earth and Mars: the \textit{Lithopanspermia} (Mars to Earth or viceversa) and the origin from a unique progenitor, that for Earth is called LUCA (the \textit{LUCA hypothesis}). We note that in the second case, independent origin for both planets, eventually present Martian lifeforms may be so different from terrestrial ones from a biochemical  point of view that it could hardly be detected with standard biochemical searches. 

To test the possibility that some lifeforms similar to the terrestrial ones may survive on Mars, we designed and built two simulators of Martian environments where to perform experiments with different bacterial strains. The biggest one is named LISA (Italian acronym for \textit{Italian laboratory for environment simulation}) and allows six simultaneous experiments. The other is a single-experiment version of it, called mini-LISA. Our LISA environmental chambers can reproduce the conditions of many Martian locations near the surface trough changes of temperature, pressure, UV fluence and atmospheric composition. Both simulators are open to collaboration with other laboratories interested in performing experiments on many kind of samples (biological, minerals, electronic) in situations similar to that of the red planet. 
 
Inside LISA we have studied the survival of several bacterial strains and endospores. We verified that the UV light is the major responsible of cell death. Neither the low temperature, nor the pressure, nor the desiccation or the atmospheric changes were effective in this sense. As expectd from previous studies, we found that some \textit{Bacillus} strains have a particular capability to survive for some hours in Martian conditions without being screened by dust or other shields.   
 
We also simulated the coverage happening on a planet by dust transported by the winds, blowing on the samples a very small quantity of volcanic ash grains or red iron oxide particles. Samples covered by these dust grains have shown a high percentage of survival, confirming that under the surface dust, if life were to be present on Mars in the past, some bacteria colonies or cells could still be present.  

\end{abstract}


\end{frontmatter}


\section{Introduction}
\label{Intro}
Life on Earth has been found in several harsh environments: near volcanic chimney at very high temperature, in rocks at high pressure or in iced conditions. When in absence of liquid water, lifeforms assume a particular state that allows them to suspend their metabolism until more suitable conditions are restored. For bacterial strains this happen by building endospores, dormant non-reproductive structure containing a thick internal wall that encloses its cell DNA and part of its cytoplasm. The structure of most endospores ensure the survival of the bacterium through periods of environmental stress and inside it most part of the water is removed. It is know that the freezing or the rapid (explosive) sublimation of the water at very low pressure is the cause of vegetative cell death \citep{Nussinov1983}. So, the pressure and temperature values allowing the existence of liquid water are compulsory to produce a biological activity, characterized by production or absorption of gases and proteins. On the contrary, very cold and dry environments or vacuum allow only a stand-by condition in terrestrial microbes.

The possibility that lifeforms may adapt to an environment different from the one of Earth, even in the space, has been studied since long time ago \citep{horneck1981}. As described below, in the last years a large number of space missions carried in the circumterrestrial space bacterial and fungal spores, as well as lichens. The limit conditions for their survival have been studied and the main factors of inactivation and mutation have been analyzed \citep[see][]{horneck2007,sancho2007}. 

In addition to space experiments, many studies in simulated planetary conditions were performed since the begin of this century with various strains of \textit{Bacillus} endospores, \textit{cyanobacteria} and different types of bacteria in saltwater soils  \citep{schuerger2003,Cockell2005,Diaz2006}. Main goal of these studies was to understand the potential contamination of planets like Mars by bacteria accidentally left on spacecraft surfaces. It has been found that UV radiation is the more effective factor of damage in space and that several biological sample are very resistant to various factors, such as vacuum, low temperature, soil and atmospheric chemistry \citep[see][and references therein]{schuerger2003}. Then, the presence of a soil or dust coverage \citep{mancinelli2008,stan-lotter2009} or pits and cracks acting as niches for bacteria \citep{moores2007} may allow to hardy terrestrial microrganisms to survive for years.

Several past experiments have been performed to test the \textit{Lithopanspermia} hypothesis, that describes the viable transport, from a planet to another, of microorganisms originally present on a planetary surface and launched in space together with the fragments of a big impact. If they survive to the big acceleration of the impact and to the space travel, they may colonize another planet by falling on the surface and finding an environment suitable to their metabolism. Samples of organic compounds with molecular weights between 178 and 536 daltons \citep{bowden2009}, of bacteria and lichens \citep{fajardo2005, stoffler2006, horneck2008, fajardo2005}, of seeds \citep{levoci2009} have been tested by applying a shock pressure that in some experiments reached 78 GPa. Many of the tested samples proved to be resistant to impacts in some pressure ranges, impact angles and type of coverage (rocks, epoxy containers, etc.). These tests may induce to think that in a large number of impacts and in special but not too rare conditions, lifeforms may contaminate a planet different than the native one. So, life on Earth may have been transported elsewhere or may have arrived from the outer regions of the Solar System.

\begin{figure}
		\includegraphics[width=8.8cm]{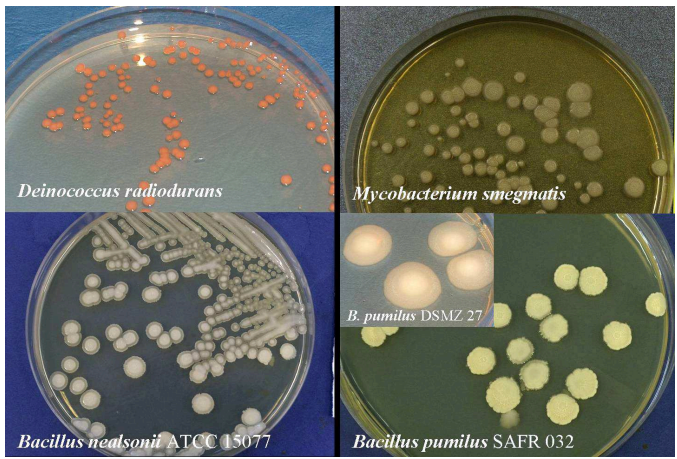}
 \caption{Colonies of various bacterial strains, cultured after exposition to Mars simulated environment. Every clear spot on a Petri dish represent a colony developed by a single cell survived to the experiment. }
 	\label{colonies}
\end{figure}

In this scenario, planet Mars represents a unique source from Panspermia to Earth or a possible case of parallel development for life. This planet had an evolution similar to Earth in the first phases of its existence. This is shown by the relict of a volcanic atmosphere and by the presence of surface morphologies resembling that of terrestrial rivers in the oldest terrains such as the Noachis Terra. If, in the past, Mars had a denser atmosphere capable to maintains liquid water on the surface, at least until the volcanic activity was able to refuel the surface by greenhouse gases, we may suppose that the same prebiotic processes happened on Earth have also happened on Mars. The end of this volcanic phase, with the consequential escape of the atmosphere and the presence of chaotic liquid flows visible in the younger terrains, may have pulled Martian lifeforms in some ecologic niches or may have extinguished them. However, the exceptional resistance of the terrestrial life in unfavorable environments described above induce the idea that Martian life may still exist. In a Lithopanspermia scenario, an impact on Mars surface may have transported a few billion years ago a life form to our planet, allowing it to survive. But it is also possible to think that similar physical and chemical conditions on both planets may have produced the same results and so that both Earth and Mars didn't need a life colonization to produce a own Biology.

The discovery of methane emission from an high latitude site on this planet \citep{krasnopolsky2004, formisano2004, formisano2008, mumma2009} reopened several astrobiological questions: if one was to discover that the origin of Martian methane is biological, this would suggest a similar mechanism to that of terrestrial methanogen bacteria. If other abiotic mechanisms such as serpentinization are responsible of this production, \citep{oze2005} Martian life may remain in the realm of the possibilities.

This is an important point: assuming that life can transit from a planet to another, one may expect that many biological mechanisms will remain similar even in a very different environment. Martian life, if discovered, should then have similar characteristics to those we already know of for life on Earth, because of a direct relation of progeny. 
On the other hand, if we assume that life arose independently on different planets, the native living beings will have, in principle, different original characteristics and different evolutionary paths. 
But we know that all lifeforms on Earth share several peculiarities: for instances, they have a unique molecular chirality (L-aminoacids coupled with D-sugars), use only twenty aminoacids to produce proteins, possess a genetic code that decides what protein is encoded by a sequence of three bases in the gene. These factors are so special and apparently not produced by any know physical and chemical reasons, to lead to think that all terrestrial lifeforms arose from one original being, called LUCA, an acronym of Last Universal Common Ancestor \citep{deduve2003}. On this basis, one may expect that an hypothetic Martian life, starting from a different ancestor, could have different genetic code, different bases, and could use different aminoacids and proteins for its metabolism. In other worlds, a potential `genuine' Martian life may remain undetected to any accurate instrument we shall send on the planet simply because it has been designed to detect byproducts or compounds that belongs to present time LUCA's progeny.

The discussion if this paper about limits of Martian life will be then dominated by this \textit{caveat}: we don't know anything about Martian life but we are obliged to use terrestrial surrogates and to reason with mechanisms such as the methane production in terms of terrestrial methanogens.      

In any case, our conclusions may be useful in the context of the panspermia or if, for still unknown reasons, the above biochemical peculiarities should appear in any lifeforms without the need of a unique progenitor. 

\begin{figure}
		\includegraphics[width=8.8cm]{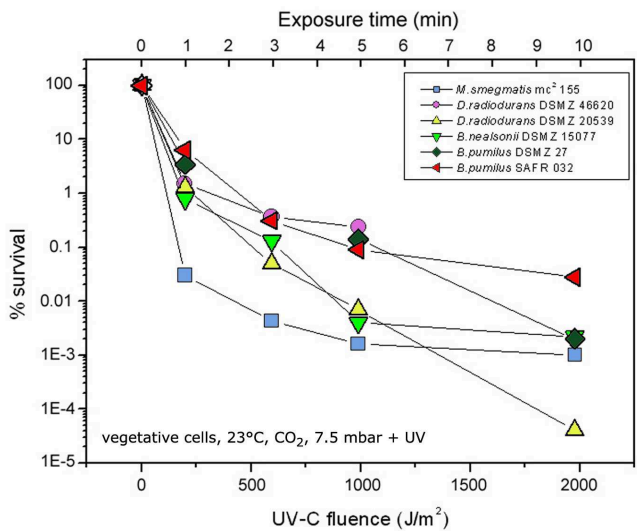}
 \caption{Survival in simulated Martian conditions for vegetative cells. all the strains are killed in a few minutes. }
 	\label{cells}
\end{figure}

\section{LISA simulators}
\label{LISA}
By keeping in mind all these limits, we designed and built with the technical support of the Center of Study and Space Activity of Padua (CISAS) a simulator of Martian environments where to perform survival experiments of terrestrial bacteria strains: LISA (Laboratorio Italiano Simulazione Ambienti) \citep{galletta2006}, that allows six simultaneous experiments in corresponding steel reaction cells of 250 cc, each containing a bacterial or soil sample. LISA environmental chambers can reproduce the conditions of many Martian locations near the surface \citep{galletta2006,visentin2009,galletta2009}. Temperature and pressure inside the simulator are modified during the experiment in the range between +20 $^\circ$C and -100 $^\circ$C whenever we need to simulate a variation between day and night or summer and winter. The pressure is kept around 0.7 kPa(=7 mbar), and an atmosphere with a composition similar the one of Mars is pumped inside the reaction cells. A UV light with a continuous spectrum with wavelengths longer than 160 nm is switched up and down to simulate the sunshine on Mars or the night. The UVC flux is similar to that expected on the planet surface, $\sim$4.3 W/m$^2$. LISA is cooled by a 500 liters liquid nitrogen reservoir filled once per week, due to the limitation imposed by the location, guested in the XII Century old Padua Castle, where the Astronomical Observatory has its offices and laboratories. This reservoir allow experiments with a maximum length of 25 hours. To overcome this limit, we conceived a smaller, single experiment simulator, called mini-LISA \citep[described in][]{galletta2010}, where two samples are kept in a single reaction cells, piled in two aluminum dishes. Only the upper dish and the biological sample contained in it receive the UV light, while the underlying one is kept in the dark, creating a control sample. The reaction cell is closed inside a cryostat that allows, with the same reservoir, to reach 10 days of experiments or, with a weekly refueling, an indefinite length in time.     

Our simulators are useful to understand what kind of lifeforms may survive, in the form of desiccated cells or spores, on the surface or in its subsurface materials. The experiments are very less expensive than a planetary mission and may investigate on a large variety of bacterial strains or microscopic animals and plants possessing high resistance to adverse conditions. Finding lifeforms that survive in some simulated Martian situation may have a double meaning: increasing the hope to find extraterrestrial life and defining the limits for a terrestrial contamination of planet Mars.   

\begin{figure*}
		\includegraphics[width=8.8cm]{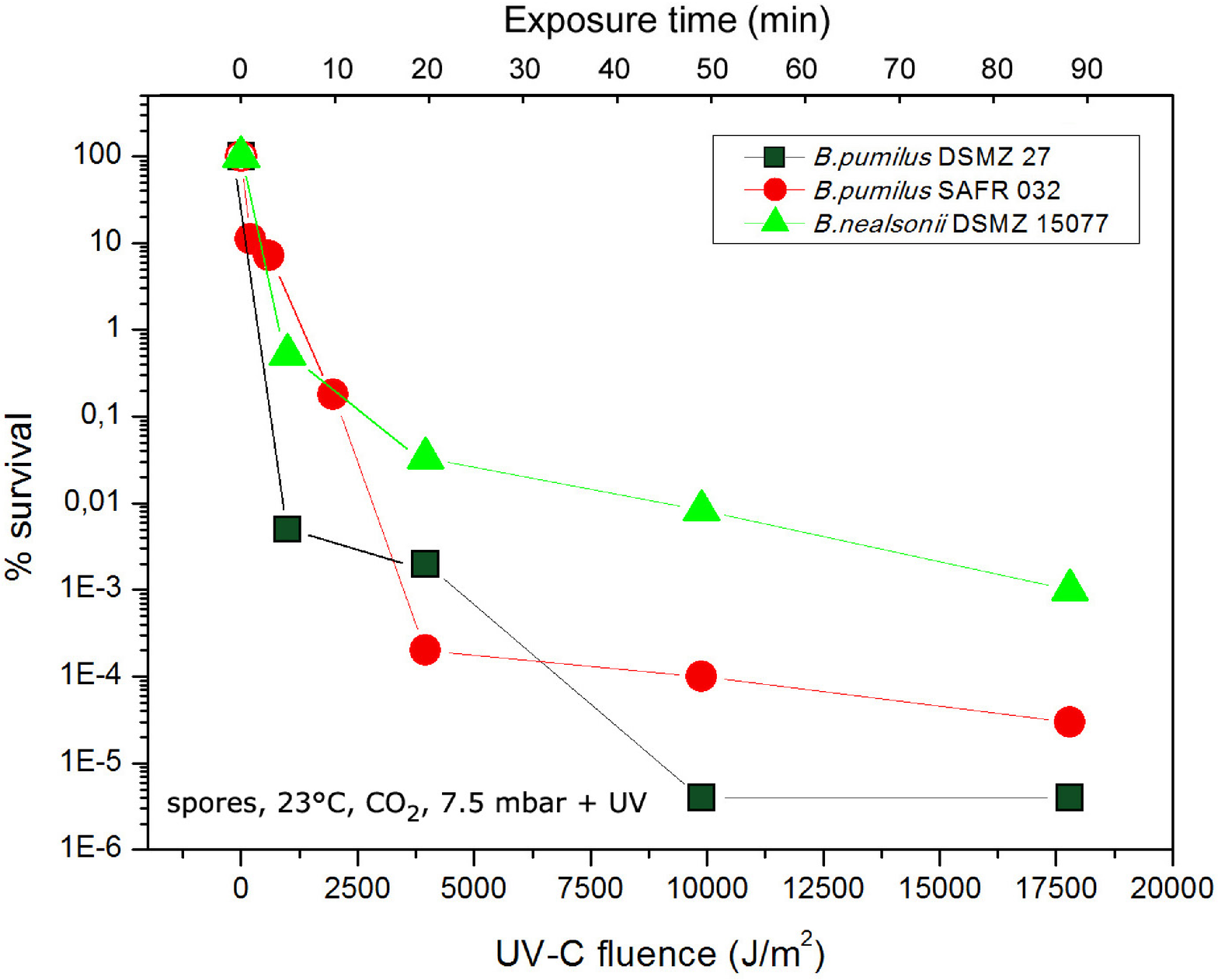}
		\includegraphics[width=8.8cm]{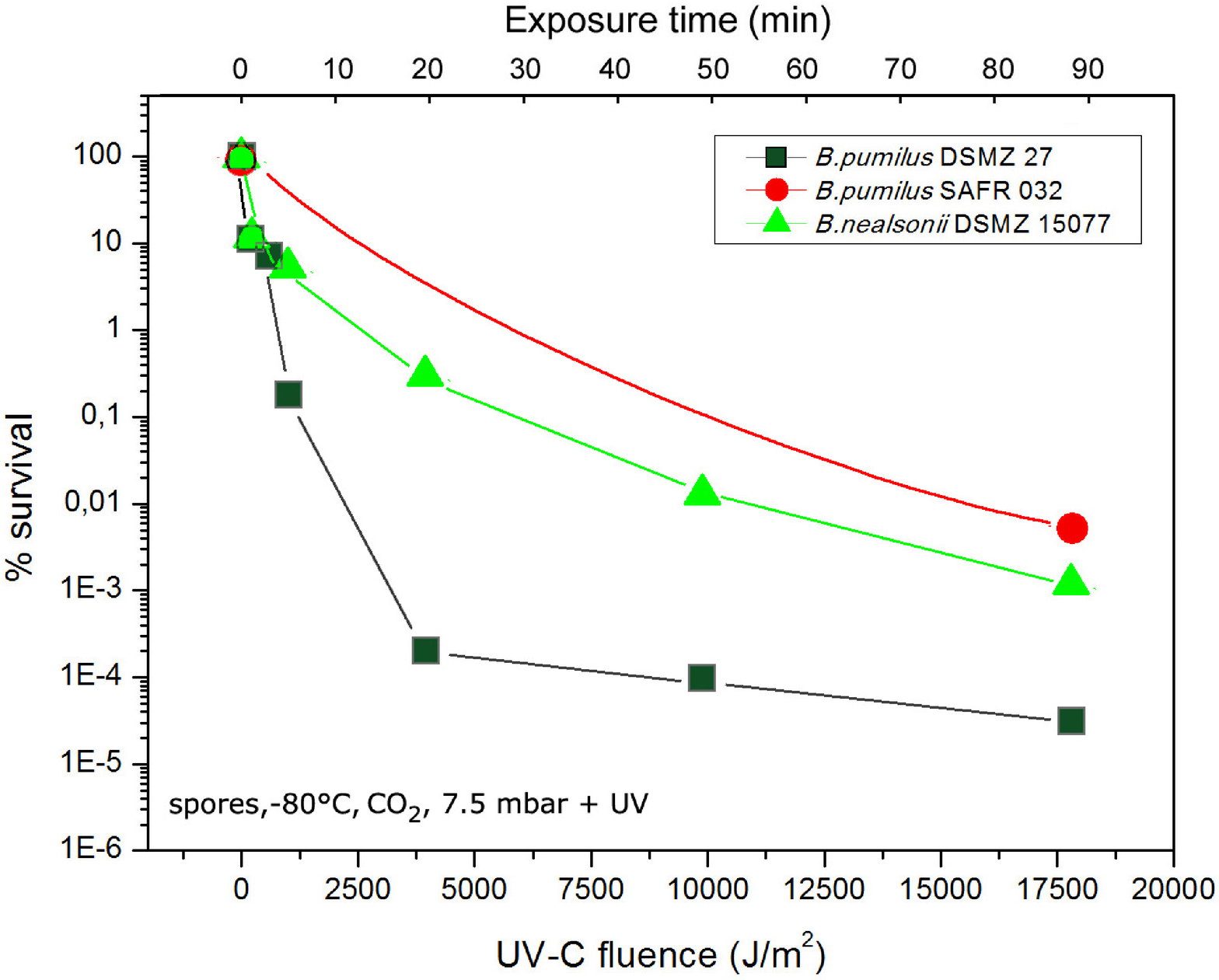}
 \caption{Survival in simulated Martian conditions for spores, much more resistant than cells to UV light and low pressure. Data are for ``summer'' (+23$^\circ$C, left panel) and ``winter'' Martian temperature(-80$^\circ$C, right panel). Note as the spores survival increase at lower temperature. }
 	\label{cold}
\end{figure*}

\section{Biological experiments}
\label{bio}
LISA and mini-LISA are two facilities left to the astro-biological community to perform experiments. We currently are running experiments for studying the survival of vegetative cells and spores, performed at the local Department of Histology, Microbiology and Medical Biotechnologies, but also experiment with samples sent to us by other Italian and foreign laboratories. In our experiments, cellular or endospores suspensions were layered on sterile coverslip dehydrated under sterile air flux, introduced in dedicated plates, closed in the steel reaction cells and then exposed to the Martian condition. When returned in the biological laboratory, the samples were extracted from the reaction cells, re-idratated and then plated in duplicate on solid media. If colonies are formed (see Figure \ref{colonies}), they are counted to establish the survival percentage with respect to the original sample.  

In general, as visible in Figure \ref{cells}, vegetative cells at laboratory temperature are soon killed by the UV radiation of the lamp simulating the Sun on Mars surface. In particular, \textit{Deinococcus radiodurans}, a bacterium known to resist to an instantaneous dose of up to 5,000 Gy\footnote{1 Gy (=Grey) is the absorption of 1 J of energy by 1 kg of matter} of ionizing radiation with no loss of viability (while 10 Gy can kill a human) appears to not resist to the combined action of the desiccation due to the low pressure (0.7 kPa) and the UV light, showing no survival in our samples after 5-8 minute of exposure.
  
A higher resistance is shown by the \textit{Bacillus pumilus} SAFR032 strain\footnote{kindly provided by prof. K.Venkateswaran, JPL, CA, USA}, whose survival is still 10\% after 10 minutes exposure. However, no cells survive after a few minutes of irradiation. Cooling of these cells produce an additional distruption, due to the sublimation of the water present inside the cells or to its conversion to ice.  This means that, as expected, no biological activity of vegetative cells, rich in water for more than 75\%, is possible in Martian conditions. 

A greater resistance is shown by spores. Two strains of \textit{Bacillus pumilus} and one of \textit{Bacillus nealsonii}, after an initial fast decrease tend to stabilize the number of `survivors' (Figure \ref{cold}, left panel).  In this case, even after a 1.5 hours experiment, a significant number of colonies is formed when the spores are plated on solid media.

Then we have freezed the samples, bringing them to winter day temperature on Mars (typically -80 $^\circ$C) and exposing them for longer time to UV radiation. These experiments are still ongoing, but preliminary results indicates that one of our \textit{Bacillus} strains has a particular capability to survive in Martian conditions without screening by dust or other shields (see the right panel of Figure \ref{cold} in comparison with the left one. As endospores suspension, it survive at least 4 hours and in one case up to 28 hours of Martian conditions. 

We found that UV light is the only mechanism able to kill the bacteria, being not effective in this sense neither the low temperature or the pressure, nor the desiccation or the atmospheric changes. When UV light is removed (on Mars this may happen during the night or in the perennial shadow of a rock or under the permafrost), spores appears to remain in a stable status, and are activated again when liquid water is added. 

We simulated the dust coverage happening on the planet by blowing on the samples very small quantity of grains of volcanic ash or red iron oxide dust used in industrial color production. Volcanic ash used was generated during a Etna Volcano eruptions and collected in the northern coast of Sicily, about 70 km away from the erupting crater. This ensure that the ash is particularly thin, such as the dust moved by the wind on the red planet. In some experiments, it has been mixed with red iron oxide. The samples covered by these dust grains have shown a high percentage of survival, indicating that under the surface dust, if life was present on Mars in the past, some bacteria colonies may still be present.  \\

\textbf{Acknowledgements} \\
The authors thanks the Air Liquide Italy for kindly supporting the LISA simulators with liquid nitrogen supplies. This work has been supported by the research funds (ex 60\%) of Padua University.


\bibliographystyle{elsarticle-harv}
\bibliography{Galletta_etal_2010_2columns}

\begin{thebibliography}{25}
\expandafter\ifx\csname natexlab\endcsname\relax\def\natexlab#1{#1}\fi
\expandafter\ifx\csname url\endcsname\relax
  \def\url#1{\texttt{#1}}\fi
\expandafter\ifx\csname urlprefix\endcsname\relax\def\urlprefix{URL }\fi

\bibitem[{{Bowden} et~al.(2009){Bowden}, {Parnell}, and
  {Burchell}}]{bowden2009}
{Bowden}, S.~A., {Parnell}, J., {Burchell}, M.~J., Jan. 2009. {Survival of
  organic compounds in ejecta from hypervelocity impacts on ice}. International
  Journal of Astrobiology 8, 19--25.

\bibitem[{{Cockell} et~al.(2005){Cockell}, {Schuerger}, {Billi}, {Imre
  Friedmann}, and {Panitz}}]{Cockell2005}
{Cockell}, C.~S., {Schuerger}, A.~C., {Billi}, D., {Imre Friedmann}, E.,
  {Panitz}, C., Jun. 2005. {Effects of a Simulated Martian UV Flux on the
  Cyanobacterium, Chroococcidiopsis sp. 029}. Astrobiology 5, 127--140.

\bibitem[{{de Duve}(2003)}]{deduve2003}
{de Duve}, C., Dec. 2003. {A Research Proposal on the Origin Of Life. Closing
  Lecture given at the ISSOL Congress in Oaxaca, Mexico, on July 4, 2002}.
  Origins of Life and Evolution of the Biosphere 33, 559--574.

\bibitem[{{Diaz} and {Schulze-Makuch}(2006)}]{Diaz2006}
{Diaz}, B., {Schulze-Makuch}, D., Apr. 2006. {Microbial Survival Rates of
  Escherichia coli and Deinococcus radiodurans Under Low Temperature, Low
  Pressure, and UV-Irradiation Conditions, and Their Relevance to Possible
  Martian Life}. Astrobiology 6, 332--347.

\bibitem[{{Fajardo-Cavazos} et~al.(2005){Fajardo-Cavazos}, {Link}, {Melosh},
  and {Nicholson}}]{fajardo2005}
{Fajardo-Cavazos}, P., {Link}, L., {Melosh}, H.~J., {Nicholson}, W.~L., Dec.
  2005. {Bacillus subtilis Spores on Artificial Meteorites Survive
  Hypervelocity Atmospheric Entry: Implications for Lithopanspermia}.
  Astrobiology 5, 726--736.

\bibitem[{{Fendrihan} et~al.(2009){Fendrihan}, {B{\'e}rces}, {Lammer}, {Musso},
  {Ront{\'o}}, {Polacsek}, {Holzinger}, {Kolb}, and
  {Stan-Lotter}}]{stan-lotter2009}
{Fendrihan}, S., {B{\'e}rces}, A., {Lammer}, H., {Musso}, M., {Ront{\'o}}, G.,
  {Polacsek}, T.~K., {Holzinger}, A., {Kolb}, C., {Stan-Lotter}, H., Feb. 2009.
  {Investigating the Effects of Simulated Martian Ultraviolet Radiation on
  Halococcus dombrowskii and Other Extremely Halophilic Archaebacteria}.
  Astrobiology 9, 104--112.

\bibitem[{{Formisano} et~al.(2004){Formisano}, {Atreya}, {Encrenaz},
  {Ignatiev}, and {Giuranna}}]{formisano2004}
{Formisano}, V., {Atreya}, S., {Encrenaz}, T., {Ignatiev}, N., {Giuranna}, M.,
  Dec. 2004. {Detection of Methane in the Atmosphere of Mars}. Science 306,
  1758--1761.

\bibitem[{{Galletta} et~al.(2010){Galletta}, {Bertoloni}, and
  {D'Alessandro}}]{galletta2010}
{Galletta}, G., {Bertoloni}, G., {D'Alessandro}, M., Mar. 2010. {New approaches
  to the exploration: planet Mars and bacterial life.} In: IAU Symposium. Vol.
  269 of IAU Symposium. pp. 0--0.

\bibitem[{{Galletta} et~al.(2009){Galletta}, {D'Alessandro}, {Bertoloni},
  {Castellani}, and {Visentin}}]{galletta2009}
{Galletta}, G., {D'Alessandro}, M., {Bertoloni}, G., {Castellani}, F.,
  {Visentin}, R., Oct. 2009. {Surviving on Mars: test with LISA simulator}.
  ArXiv e-prints.

\bibitem[{{Galletta} et~al.(2006){Galletta}, {Ferri}, {Fanti}, {D'Alessandro},
  {Bertoloni}, {Pavarin}, {Bettanini}, {Cozza}, {Pretto}, {Bianchini}, and
  {Debei}}]{galletta2006}
{Galletta}, G., {Ferri}, F., {Fanti}, G., {D'Alessandro}, M., {Bertoloni}, G.,
  {Pavarin}, D., {Bettanini}, C., {Cozza}, P., {Pretto}, P., {Bianchini}, G.,
  {Debei}, S., Dec. 2006. {S.A.M., the Italian Martian Simulation Chamber}.
  Origins of Life and Evolution of the Biosphere 36, 625--627.

\bibitem[{{Geminale} et~al.(2008){Geminale}, {Formisano}, and
  {Giuranna}}]{formisano2008}
{Geminale}, A., {Formisano}, V., {Giuranna}, M., Jul. 2008. {Methane in Martian
  atmosphere: Average spatial, diurnal, and seasonal behaviour}. Planetary and
  Space Science 56, 1194--1203.

\bibitem[{{Horneck}(1981)}]{horneck1981}
{Horneck}, G., 1981. {Survival of microorganisms in space: a review}. Advances
  in Space Research 1, 39--48.

\bibitem[{{Horneck} and {Rabbow}(2007)}]{horneck2007}
{Horneck}, G., {Rabbow}, E., 2007. {Mutagenesis by outer space parameters other
  than cosmic rays}. Advances in Space Research 40, 445--454.

\bibitem[{{Horneck} et~al.(2008){Horneck}, {St{\"o}ffler}, {Ott}, {Hornemann},
  {Cockell}, {Moeller}, {Meyer}, {de Vera}, {Fritz}, {Schade}, and
  {Artemieva}}]{horneck2008}
{Horneck}, G., {St{\"o}ffler}, D., {Ott}, S., {Hornemann}, U., {Cockell},
  C.~S., {Moeller}, R., {Meyer}, C., {de Vera}, J., {Fritz}, J., {Schade}, S.,
  {Artemieva}, N.~A., Feb. 2008. {Microbial Rock Inhabitants Survive
  Hypervelocity Impacts on Mars-Like Host Planets: First Phase of
  Lithopanspermia Experimentally Tested}. Astrobiology 8, 17--44.

\bibitem[{{Krasnopolsky} et~al.(2004){Krasnopolsky}, {Maillard}, and
  {Owen}}]{krasnopolsky2004}
{Krasnopolsky}, V.~A., {Maillard}, J.~P., {Owen}, T.~C., Nov. 2004. {First
  Detection of Methane in the Martian Atmosphere: Evidence for Life?} In:
  Bulletin of the American Astronomical Society. Vol.~36 of Bulletin of the
  American Astronomical Society. pp. 1127--+.

\bibitem[{{Levoci} et~al.(2009){Levoci}, {Burchell}, and {Tepfer}}]{levoci2009}
{Levoci}, G., {Burchell}, M.~J., {Tepfer}, D., Mar. 2009. {Survival of Seeds in
  Impacts at 1 km s-1 and Above}. In: Lunar and Planetary Institute Science
  Conference Abstracts. Vol.~40 of Lunar and Planetary Institute Science
  Conference Abstracts. pp. 1239--+.

\bibitem[{{Moores} et~al.(2007){Moores}, {Smith}, {Tanner}, {Schuerger}, and
  {Venkateswaran}}]{moores2007}
{Moores}, J.~E., {Smith}, P.~H., {Tanner}, R., {Schuerger}, A.~C.,
  {Venkateswaran}, K.~J., Dec. 2007. {The shielding effect of small-scale
  martian surface geometry on ultraviolet flux}. Icarus 192, 417--433.

\bibitem[{{Mumma} et~al.(2009){Mumma}, {Villanueva}, {Novak}, {Hewagama},
  {Bonev}, {DiSanti}, {Mandell}, and {Smith}}]{mumma2009}
{Mumma}, M.~J., {Villanueva}, G.~L., {Novak}, R.~E., {Hewagama}, T., {Bonev},
  B.~P., {DiSanti}, M.~A., {Mandell}, A.~M., {Smith}, M.~D., Feb. 2009. {Strong
  Release of Methane on Mars in Northern Summer 2003}. Science 323, 1041--.

\bibitem[{{Nussinov} and {Lysenko}(1983)}]{Nussinov1983}
{Nussinov}, M.~D., {Lysenko}, S.~V., Dec. 1983. {Cosmic Vacuum Prevents
  Radiopanspermia}. Origins of Life 13, 153--164.

\bibitem[{{Osman} et~al.(2008){Osman}, {Peeters}, {La Duc}, {Mancinelli},
  {Ehrenfreund}, and {Venkateswaran}}]{mancinelli2008}
{Osman}, S., {Peeters}, Z., {La Duc}, M.~T., {Mancinelli}, R., {Ehrenfreund},
  P., {Venkateswaran}, K., Feb. 2008. {Effect of Shadowing on Survival of
  Bacteria under Conditions Simulating the Martian Atmosphere and UV
  Radiation}. pplied and Environmental Microbiology 74, 959--970.

\bibitem[{{Oze} and {Sharma}(2005)}]{oze2005}
{Oze}, C., {Sharma}, M., May 2005. {Have olivine, will gas: Serpentinization
  and the abiogenic production of methane on Mars}. Geophysical Research
  Letters 32, 10203--+.

\bibitem[{{Sancho} et~al.(2007){Sancho}, {de la Torre}, {Horneck}, {Ascaso},
  {de los Rios}, {Pintado}, {Wierzchos}, and {Schuster}}]{sancho2007}
{Sancho}, L.~G., {de la Torre}, R., {Horneck}, G., {Ascaso}, C., {de los Rios},
  A., {Pintado}, A., {Wierzchos}, J., {Schuster}, M., Jun. 2007. {Lichens
  Survive in Space: Results from the 2005 LICHENS Experiment}. Astrobiology 7,
  443--454.

\bibitem[{{Schuerger} et~al.(2003){Schuerger}, {Mancinelli}, {Kern},
  {Rothschild}, and {McKay}}]{schuerger2003}
{Schuerger}, A.~C., {Mancinelli}, R.~L., {Kern}, R.~G., {Rothschild}, L.~J.,
  {McKay}, C.~P., Oct. 2003. {Survival of endospores of Bacillus subtilis on
  spacecraft surfaces under simulated martian environments:implications for the
  forward contamination of Mars}. Icarus 165, 253--276.

\bibitem[{{St{\"o}ffler} et~al.(2006){St{\"o}ffler}, {Meyer}, {Fritz},
  {Horneck}, {M{\"o}ller}, {Cockell}, {Ott}, {de Vera}, {Hornemann}, and
  {Artemieva}}]{stoffler2006}
{St{\"o}ffler}, D., {Meyer}, C., {Fritz}, J., {Horneck}, G., {M{\"o}ller}, R.,
  {Cockell}, C.~S., {Ott}, S., {de Vera}, J.~P., {Hornemann}, U., {Artemieva},
  N.~A., Mar. 2006. {Impact Experiments in Support of ''Lithopanspermia'': The
  Route from Mars to Earth}. In: {S.~Mackwell \& E.~Stansbery} (Ed.), 37th
  Annual Lunar and Planetary Science Conference. Vol.~37 of Lunar and Planetary
  Institute Science Conference Abstracts. pp. 1551--+.

\bibitem[{{Visentin} et~al.(2009){Visentin}, {Bertoloni}, {D'Alessandro}, and
  {Galletta}}]{visentin2009}
{Visentin}, R., {Bertoloni}, G., {D'Alessandro}, M., {Galletta}, G., Aug. 2009.
  {in: Abstracts from the 2008 Issol Meeting}. Origins of Life and Evolution of
  the Biosphere 39, 179--392.

\end{thebibliography}





\end{document}